\documentclass[onecollarge,natbib]{svjour2}
\bibpunct{[}{]}{,}{n}{}{,} 
\smartqed  
\usepackage{graphicx}
\newcommand{\beqn}{\begin{equation}}
\newcommand{\eeqn}{\end{equation}}
\newcommand{\barr}[1]{\begin{array}{#1}}
\newcommand{\earr}{\end{array}}
\newcommand{\beqna}{\begin{eqnarray}}
\newcommand{\eeqna}{\end{eqnarray}}

\journalname{Few-Body Systems (APFB2011)}
\begin{document}

\title{\boldmath
Proton strangeness form factors in (4,1) clustering configurations
\thanks{Presented at The Fifth Asia-Pacific Conference on Few-Body Problems in Physics 2011 in Seoul, South Korea, 22-26 August 2011.
This work is supported in part by the National Science Council of the Republic of China (Taiwan) under grant No. NSC99-2112-M002-011 and Center for Theoretical Sciences,
National Taiwan University.}
}
\subtitle{}


\author{Alvin Kiswandhi        \and
  Hao-Chun Lee \and
  Shin Nan Yang
}


\institute{Alvin Kiswandhi \and  Hao-Chun Lee \and Shin Nan Yang\at
  Department of Physics and Center for Theoretical Sciences, National Taiwan University, Taipei 106, Taiwan R.O.C.\\
              \email{alvin@phys.ntu.edu.tw}           
}

\date{Received: date / Accepted: date}

\maketitle

\begin{abstract}
We reexamine a recent result within a nonrelativistic constituent quark model (NRCQM) which maintains that the $uuds\bar s$ component in the proton
has its $uuds$ subsystem in $P$ state, with its $\bar s$ in $S$ state (configuration I). When the result are corrected, contrary to the previous result,
we find that all the empirical signs of the form factors data can be described by the lowest-lying $uuds\bar s$ configuration with $\bar s$ in $P$ state that
has its $uuds$ subsystem in $S$ state (configuration II). Further, it is also found that the
removal of the center-of-mass (CM) motion of the clusters will
enhance the contributions of the transition current considerably. We also show that
a reasonable description of the existing
form factors data can be obtained with a very small probability $P_{s\bar s}=0.025\%$ for
the $uuds\bar s$ component. We further see that the agreement of our prediction with the data for $G_A^s$ at low-$q^2$
region can be markedly improved by a small admixture of configuration I. It is also found that by not removing CM
motion, $P_{s\bar s}$ would be overestimated by about a factor of
four in the case when transition dominates over direct currents. Then, we also study
the consequence of a recent estimate reached from analyzing the existing
data on quark distributions that $P_{s\bar s}$ lies between $2.4-2.9\%$ which would lead to a large size for the five-quark ($5q$) system, as well as a small
bump in both $G^s_E+\eta G^s_M$ and $G^s_E$ in the region of $q^2\le 0.1 \,\textrm{GeV}^2$.

\keywords{Proton strangeness form factors \and NRCQM \and Five-quark configurations}
\end{abstract}

\section{Introduction}
\label{intro}

The proton has been widely viewed as a system consisting of three $uud$ quarks.
However, there are indications of possible existence of strangeness content
in the proton \cite{Beck01}. Later, many other efforts are suggested, including those in Refs. \cite{Amsler98,TOY} as well as the ongoing effort in $\phi$
photoproduction which is now being pursued at SPring-8 \cite{Ohta11}. Meanwhile, four parity-violating $ep$ scattering
experimental programs SAMPLE \cite{SAMPLE}, HAPPEx \cite{HAPPEX}, A4
\cite{A4}, and G0 \cite{G0} have already been successful in extracting the proton strangeness electromagnetic form factors.

On the theoretical side, lattice QCD remains the only reliable first-principle theoretical
method which could determine the strangeness form factors. For example, a recent low-mass
quenched lattice QCD simulation gives
$\mu_S=(-0.046\pm 0.019)\mu_N$ \cite{Leinweber05} and $G_E^s(Q^2=0.1~
\textrm{GeV}^2)=-0.009\pm 0.005\pm 0.003\pm 0.027$ \cite{Leinweber06}. More
recent LQCD efforts can be found in Ref. \cite{LQCD10}. Still, a study of this interesting question within NRCQM
could provide some hints concerning the underlying quark structure.

\section{Calculations and results of strangeness form factors with CM motion removed}
\label{sec:1}



The calculation of Refs. \cite{ZR05,RZ06,An06} did not remove the CM motion of the quark clusters which could affect the final results.
Accordingly, we reexamine the problem with the removal of the CM motion of the clusters and obtain
results which differ substantially from those presented in Refs. \cite{ZR05,RZ06,An06}.

The configurations of the $uuds\bar s$ component in the proton
considered in Refs. \cite{ZR05,RZ06,An06} are all of (4,1) clustering
type in that either four quarks $uuds$ would be in $P$ state with $\bar
s$ in $S$ state (configuration I) or $uuds$ in $S$ state while $\bar s$
in $P$ state (configuration II), respectively. After the degeneracy is lifted by the color hyperfine
quark-quark interaction as shown in Ref. \cite{ZR05}, the states of the lowest energy in configurations I and
II for $uuds$ cluster would have the space, flavor, and spin state symmetry
of $[31]_X [4]_{FS}[22]_F[22]_S$ and $[4]_X[31]_{FS}[211]_F[22]_S$, respectively
\cite{ZR05}. We will focus only on these two states of the lowest
energy in this study.

In the case of configuration I, after the CM motion of the $5q$ cluster is
removed, we obtain the following results for the the contributions of the diagonal ($D$) and non-diagonal $(\textit{ND})$ matrix
elements of the current to the proton strangeness form factors,
\beqna
G_E^{s,D}(q^2) &=& - \frac{q^2}{24\omega_5^2} e^{-q^2/5\omega_5^2} P_D;\quad G_E^{s,\scriptsize{\textit{ND}}}(q^2) = \delta C_{35}^{2/3}\frac{2\cdot15^{3/4}}{9\sqrt{3}}\frac{q^2}{m_s\omega_5} e^{-4q^2/15\omega_5^2}P_{\scriptsize{\textit{ND}}}\,\label{IGEwoCM} \\
G_M^{s,D}(q^2) &=&  \frac{m_p}{2m_s} e^{-q^2/5\omega_5^2} P_D; \quad G_M^{s,\scriptsize{\textit{ND}}}(q^2) = \delta C_{35}^{2/3}\frac{2\cdot15^{3/4}}{9\sqrt{3}}\frac{4m_p}{\omega_5} e^{-4q^2/15\omega_5^2}P_{\scriptsize{\textit{ND}}}\,,\label{IGMwoCM}  \\
G_A^{s,D}(q^2) &=& -\ \frac{1}{3} e^{-q^2/5\omega_5^2} P_D; \quad G_A^{s,\scriptsize{\textit{ND}}}(q^2) = \delta C_{35}^{2/3}\frac{2\cdot15^{3/4}}{9\sqrt{3}}\frac{3\omega_5}{m_s}e^{-4q^2/15\omega_5^2}P_{\scriptsize{\textit{ND}}}\,\label{IGAwoCM}
\eeqna
where $P_D \equiv P_{s \bar s}$ and $P_{\scriptsize{\textit{ND}}} \equiv \sqrt{ P_{uud}  P_{s\bar s}}$. Also, $\omega_3$, $P_{uud}$ and $\omega_5$,
$P_{s\bar{s}}$ denote the usual oscillator parameters and probabilities, respectively, of the $uud$ and $uuds\bar s$
configurations in the proton, and $C_{35}\equiv[2\omega_3\omega_5/(\omega_3^2+\omega_5^2)]^{9/2}$,
while $\delta$ denotes the relative phase between the $uud$ and $uuds\bar s$ components of the wave functions in the proton. As in the case before the removal of the CM motion, both $G_M^{s,ND}$ and $G_A^{s,ND}$ are of the same sign. Consequently, as long as the transition current contributions dominates, the configuration with $\bar s$ in $S$ state cannot be the dominant configuration for $uuds\bar s$ component, contrary to the findings of Ref. \cite{RZ06}.


In configuration II, $uuds$ cluster is in $S$ state while $\bar s$ is in $P$ state. The results with the
removal of CM motion will be presented are
\beqna
G_E^{s,D} &=& \frac{q^2}{8\omega_5^2} e^{-q^2/5\omega_5^2} P_D;\quad G_E^{s,\scriptsize{\textit{ND}}} = \delta C_{35}^{2/3}\sqrt{\frac{2}{5}}\left(\frac{5}{3}\right)^{3/4}\frac{q^2}{m_s\omega_5}e^{-4q^2/15\omega_5^2} P_{\scriptsize{\textit{ND}}}, \label{IIGEwoCM}\\
G_M^{s,D} &=& \frac{m_p}{m_s} \left(\frac{-1}{6} - \frac{2q^2}{15\omega_5^2}\right) e^{-q^2/5\omega_5^2} P_D;\quad G_M^{s,\scriptsize{\textit{ND}}} = \delta C_{35}^{2/3}\sqrt{\frac{2}{5}}\left(\frac{5}{3}\right)^{3/4}\frac{4m_p}{\omega_5}e^{-4q^2/15\omega_5^2} P_{\scriptsize{\textit{ND}}},\label{IIGMwoCM} \\
G_A^{s,D} &=& \left(\frac{-1}{3} +
\frac{2q^2}{15\omega_5^2}\right) e^{-q^2/5\omega_5^2} P_D; \quad G_A^{s,\scriptsize{\textit{ND}}} = -\delta C_{35}^{2/3}\sqrt{\frac{2}{5}}\left(\frac{5}{3}\right)^{3/4}\frac{5\omega_5}{m_s}e^{-4q^2/15\omega_5^2} P_{\scriptsize{\textit{ND}}}.
\label{IIGAwoCM}
\eeqna

Here, the transition current contributions  to $G_M^s$ and $G_A^s$, as given in Eqs.
(\ref{IIGMwoCM}-\ref{IIGAwoCM}) are of opposite sign and since the transition current contributions dominate over the direct
current contributions in the model considered here, this configuration is in agreement with the data.


We take the proton and quark masses to be 0.938 and 0.313 GeV,
respectively and $\omega_3 = 0.246$ GeV. We then vary $\omega_5$ and $P_{s\bar{s}}$ to fit the experimental data $G_E^{s} + \eta G_M^{s}$
\cite{G0}, which are more directly measured in the experiments and
$G_A^{s}$ as extracted in Ref. \cite{Pate08}. Both signs of $\delta=\pm 1$
are tried and the best results are then determined.

\begin{figure}
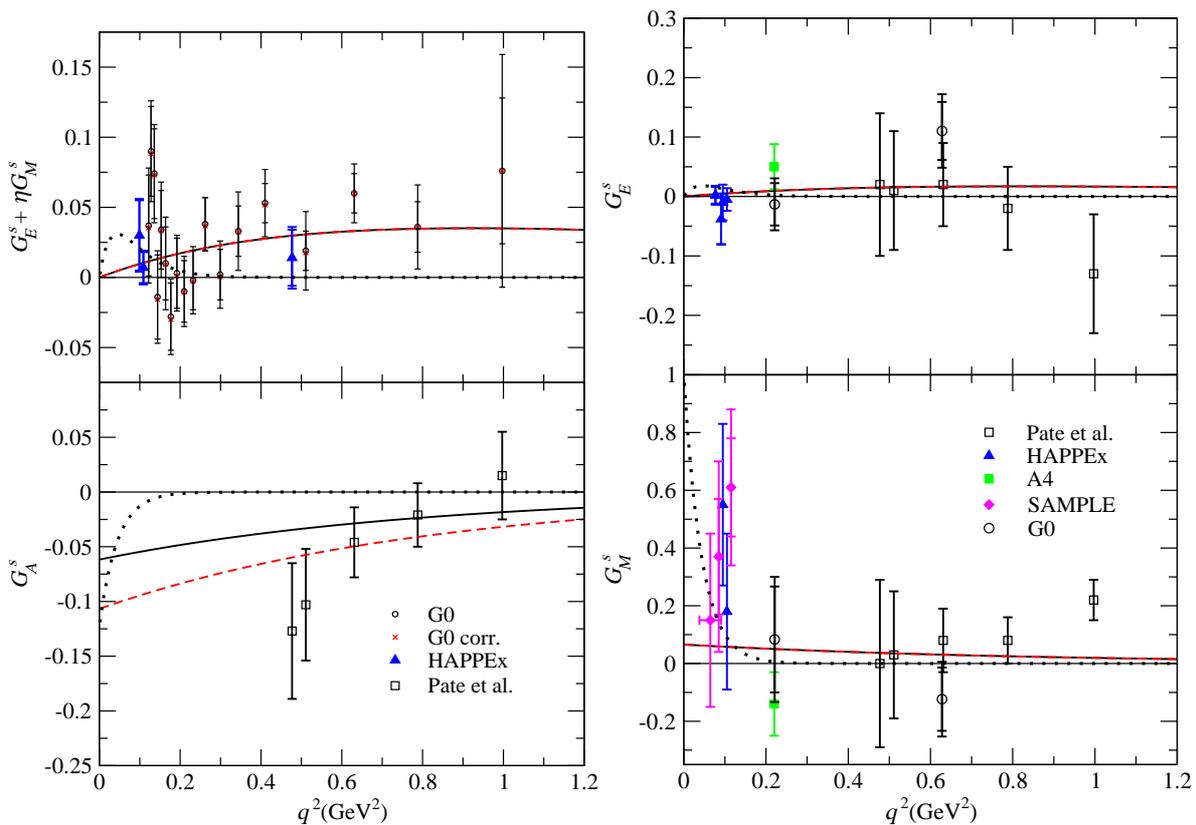

\begin{center}
\includegraphics[width=0.5\linewidth,angle=0]{figure1.eps}\includegraphics[width=0.5\linewidth,angle=0]{figure2.eps}
\end{center}
\caption{Left: Our predictions for form factors $G_E^s +\eta G_M^s$ and
$G_A^s$. The crosses are the corrected values of
the G0 data by taking into account the two-boson exchange mechanism
\cite{Zhou07}. Right: Our results for $G_E^s$ and $G_M^s$.
The results obtained with configurations with $\bar s$ in
pure $P$ state,  $S$ and $P$ states admixture A, and B are denoted,
respectively, by full, dashed, and dotted lines. Experimental data
from Refs. \cite{SAMPLE,HAPPEX,A4,G0,Pate08} are denoted by solid diamonds (SAMPLE), solid triangles (HAPPEx), solid boxes (A4), open circles (G0),
and open boxes (Pate {\it et al.}), respectively.}
\label{FF}       
\end{figure}

Our best fits to the experimental data $G_A^{s}$ and
$G_E^{s} + \eta G_M^{s}$ within configuration II are
shown in the left panel of Fig. \ref{FF} as solid curves, with $\omega_5 = 0.469$ GeV, $P_{s\bar s} =
0.025\%$, and $\delta_P = +1$, where subscript $P$ denotes the orbital state of $\bar s$.
The ensuing results for $G_E^s$ and $G_M^s$ are shown in the right panel of Fig. \ref{FF}. It
is seen that the agreement with the data are in general quite good
except for $G_E^{s} + \eta G_M^{s}$ and $G_M^s$ at small
values of $q^2$, where there are large experimental uncertainties.




We have also explored the possibility of mixing configurations II
and I, namely, $|\textrm{proton}\rangle=A_3|3q>
 +A_5\sum_{\alpha}\delta_\alpha b_\alpha|5q;\alpha>$
where $\alpha = S, P$ denotes the orbital state of $\bar s$.
We see that some improvements can be achieved only for $G_A^s$ at
low-$q^2$ region with a small mixing probability of $b^2_S=8\%$ for
configuration I, relatives phases $\delta_P=1, \delta_S=-1$, and a
combined probability of $P_{s\bar s}=A^2_5=0.058\%$ (called
admixture A), as shown by the dashed curves in Fig. \ref{FF}.

Notice that we could fit the data reasonably well with a rather
small probability of $uuds\bar s$ component. It is in sharp contrast to the values of
$P_{s\bar s}=10\sim 15\%$ required in Ref. \cite{RZ06}. It is interesting to note that our set of
harmonic oscillator model parameters would give rise to a size of
the $uuds\bar s$ to be about 0.5 fm, which is quite close to that
estimated by Ref. \cite{Henley92} using a proton-core-$\phi$ picture
for $5q$ system with a scaling factor $s = 1.5$.

We also explore the consequences of a recent work by Chang and Peng \cite{Chang11} which employs BHPS model \cite{BHPS80}
by fixing $P_{s\bar s} = 2.4\%$ and varying $\omega_5$ to fit the data. The best fit we
obtain with $\omega_5 = 0.108$ GeV, which corresponds to a large
size of the $5q$ system with $r_{5q}=2.16$ fm, and a small
admixture of $S$ state with a probability of about $15\%$ (called
admixture B), are shown in Fig. \ref{FF} by dotted
lines. The most interesting feature of this fit is the appearance of
a bump in $G^s_E+\eta G^s_M$ in the very low-$q^2$ region with
$q^2\le 0.1  \, \textrm{GeV}^2$, which seems to be hinted by the G0 data but
hampered by large experimental error bars and fluctuations. It would
be worthwhile to carry out experiments in such a low-$q^2$ region if
further theoretical study would support this behavior.

\section{Summary and conclusion}

In summary, we have reinvestigated, within a NRCQM, the question of whether a $5q$
component with configuration of (4,1) clustering, can account for the
data of the proton strangeness form factors. Two configurations (I and II) of
the lowest energies are considered.

We have not been able to reproduce the results of Ref. \cite{RZ06}
which show that configuration I is the preferred dominant configuration.
When the corrected expression for $G^s_A$ in configuration I is employed, $G^s_A$ and $G^s_M$ are
of the same sign in the low-$q^2$ region which clearly contradicts all existing data.

We then study configuration II and make an effort to remove the CM motions of the clusters. We demonstrate that it is
possible to give a satisfactory description of the existing data on
the proton strangeness form factors with a very small value of $P_{s\bar s}=0.025\%$. The agreement
with $G^s_A$ data can be improved in the low-$q^2$ region by considering an
admixture of configurations I and II with a total $uuds\bar s$
probability $P_{s\bar s}$ increased to $0.058\%$ with configuration
I accounts for $8\%$ of the total.  We further find that
without removing CM motion, $P_{s\bar s}$ would be overestimated by
about a factor of four in the case when transition current dominates.
Although it is tempting to conclude that $uuds\bar s$ arrange themselves
in configuration II, we should remember that the agreement between our results and the existing
data is not perfect, to say the least. For example, recent data from
A4 at $q^2=0.22  \, \textrm{GeV}^2$ gives a negative value of
$G^s_M=-0.14\pm0.11\pm0.11$. Also, one might ask whether NRCQM is quantitatively reliable
in evaluating the contributions of transition current which is found
to be dominant in our calculation but is of a relativistic effect in
nature.

We have also explored the consequence of a recent claim
\cite{Chang11} that $P_{s\bar s}$   lies between  $2.4-2.9\%$. A small bump
in both $G^s_E+\eta G^s_M$ and $G^s_E$ in the region of $q^2\le 0.1  \, \textrm{GeV}^2$
for an admixture of configuration I and II.


\begin{acknowledgements}
We acknowledge helpful discussions we have with Drs. C.S. An,
Fatiha Benmokhtar, C.-W. Kao, and B.S. Zou. H.C. Lee
gratefully acknowledges the warm hospitality extended to her during
a brief visit to IHEP, Beijing.
\end{acknowledgements}




\end{document}